# Surface plasmon enhanced broadband spectrophotometry on black silver substrates

Zhida Xu[1, a)], Yi Chen[1] Manas R Gartia[2], Jing Jiang[1], and Gang Logan Liu[1]

[1]Micro and Nanotechnology Laboratory, Department of Electrical and Computer Engineering, University of Illinois at Urbana-Champaign, Urbana, IL 61801, USA

[2]Department of Nuclear, Plasma and Radiological Engineering, University of Illinois at Urbana-Champaign, Urbana, IL 61801, USA

[(a)]zhidaxu1@illinois.edu

**ABSTRACT**

We demonstrate surface plasmon-induced enhancements in optical imaging and spectroscopy on silver coated silicon nanocones which we call black silver substrate. The black silver substrate with dense and homogeneous nanocone forest structure is fabricated on wafer level with a mass producible nanomanufacturing method. The black silver substrate is able to efficiently trap and convert incident photons into localized plasmons in a broad wavelength range, which permits the enhancement in optical absorption from UV to NIR range by 12 times, the visible fluorescence enhancement of ~30 times and the NIR Raman scattering enhancement factor up to ~$10^8$. We show a considerable potential of the black silver substrate in high sensitivity and broadband optical sensing and imaging of chemical and biological molecules.

[a)] zhidaxu1@illinois.edu

Black silver substrate is defined here as the silver coated black silicon substrate. Black silicon is the silicon whose surface is modified to have extremely low optical reflectivity and high absorption from visible to infrared wavelength range and thus has black surface appearance[1,2]. Recently the potential of black silicon has been recognized in some applications such as highly sensitive photodiode, solar cell, super-hydrophobicity and biomedical sensing starts to be realized and produced intentionally.[3,4,5,6,7] Black silicon can be made by reactive ion etching or femtosecond laser machining.[8,9] On the other hand, the photonic enhancements on coinage metals surface (silver, gold, copper) especially nanostructured metal surface are known as surface enhanced fluorescence (SEF), surface enhanced Raman scattering (SERS) and Surface-Enhanced Infrared Absorption-Reflectance (SEIRA). [10,11,12,13] The optical enhancement at discrete wavelengths is considered to be related with surface plasmon resonance (SPR) determined by the optical constant, size and geometry of the metal surface and surrounding media.[14] Noble metals have appropriate optical constants for SPR in visible or near infrared (NIR) range which is commonly used for excitation.[14] The nanostructures further benefit the optical enhancement upon sharp tips by 'lightning-rod' effect and plasmon coupling between adjacent particles.[14]

In this paper, with a reactive ion plasma etching method[7], we reliably produce batches of nanocone structured black silicon with whole wafer scale uniformity at room temperature in a short time and in no need of photomask. The black silver device is completed after depositing a layer of silver on the black silicon. We call it black silver because it looks much darker than smooth silver films which we call smooth silver. The enhanced broadband optical absorbance and photon trapping are demonstrated by measuring and comparing the reflectance spectra in the wavelength range from 200nm to 800nm on smooth silver and uniform black silver substrates. The fluorescence enhancement is characterized by the comparison of fluorescence spectra of Rhodamine 6G (R6g) molecules deposit on black silver and smooth silver substrate. Furthermore surface enhancement Raman spectroscopy (SERS) detections of R6G and oligopeptides is demonstrated, which exhibit uniformly high enhancement factor and potential application for high sensitivity label-free sensing.

The nanocone forest structures on the black silicon substrate is produced by reactive oxygen and bromine ion mixture plasma etching, in which bromine ion plays the role of etching while oxygen ion plays the role of oxidized passivation. In this process, the aspect ratio and the etching rate of the silicon nanocones can be controlled by oxygen passivation time, flux rate, and bromine etching time.[7] With this etching-passivation process, we can reliably produce the dense and uniform nanocones all over the single crystalline silicon wafer (<100> n-type), which makes the whole wafer "black" (**Fig. 1a**) or lithographically patterned areas "black" (**Fig. 1b**). Our optical measurement indicate >99% optical absorption (not shown) in visible wavelengths by our black silicon substrate. The nanocone structure is the key to produce perfectly black silicon as this special structure provides a graded optical reflective index layer on surface to eliminate the reflection due to mismatch of dielectric constants at the material interface.

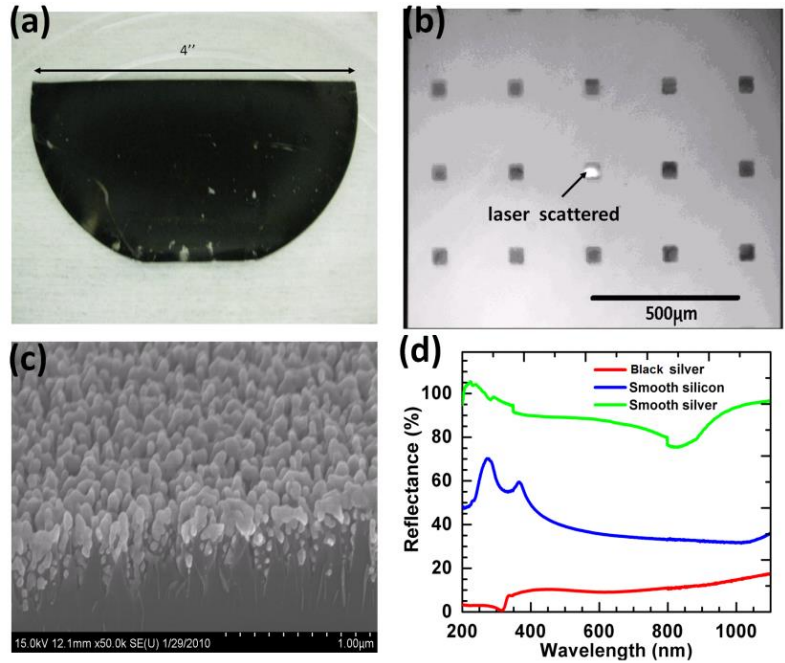

Figure 1. Black silicon and black silver. (a) Photograph of black silicon wafer (half) (b) Photograph of square array patterned black silver. The shining spot in the center square is induced by a laser focal spot (c) SEM image of the cross section of black silver substrate. (d) Reflectance spectra of black silver, smooth silver and smooth silicon wafer.

To make the black silver substrate, we deposit 5 nm thick Titanium and 80 nm thick silver directly on top of the black silicon. **Fig. 1c** shows the scanning electron microscope (SEM) image of the cross-section of the black silver substrate, in which we can see the silicon nanocone forest is covered by a layer of silver on top, especially on the tip of cones. The nanocones are around 500 nm tall, 180 nm wide at the base. The spacing between two adjacent silicon nanocones without silver coating is about 100nm and the spacing is reduced to sub-50 nm after silver coating. We observed by naked eye that while silver coated smooth silicon or the smooth silver substrate looks bright and shining like mirror, the silver coated black silicon substrate looks much darker than and not as shining as the smooth silver substrate shown as the black squares in Fig. 1b. To quantify the optical reflectance and absorbance of black silver substrates, we measured and compared the reflectance spectra (**Fig. 1d**) in the wavelength range from 200nm to 1100nm with a UV-Vis-IR optical spectrophotometer (Varian Gary 5G). The distinction between reflectance on smooth silver, black silver and smooth silicon wafer are significant. The reflectance of smooth silver is above 80% while the reflectance of black silver is below 20% in the entire wavelength range. The reflectance of smooth silicon always resides in between the former two. The averaged reflectance over all wavelength range is 92.5% for smooth silver, 51.2% for smooth silicon and 9.9% for black silver. With the assumption of no UV-Vis-NIR transmission for non-transparent substrate, the averaged absorbance of black silver is 90.1%, 12 times higher than the averaged absorbance of smooth silver (7.5%), which agrees with the reported calculation earlier[13]. Since silver has been known with low loss plasmonic resonance, we can assume that most of incident photons are trapped in the silver coated nanocone forest and converted to localized and surface plasmons, which is also indicated by the simulation result shown later. As the matter of fact the reflectance from the black silver substrate is distinctively different from the reflectance from smooth silver substrate. For smooth silver substrate, the reflectance is primary consisted of specular and diffuse reflection of incident light while for black silver substrates the measured reflectance is the scattering photon emission from the resonating plasmons on surface. The maximum absorption measured on the black silver substrate occurring around the bulk plasma wavelength of silver at 320 nm supports this point.

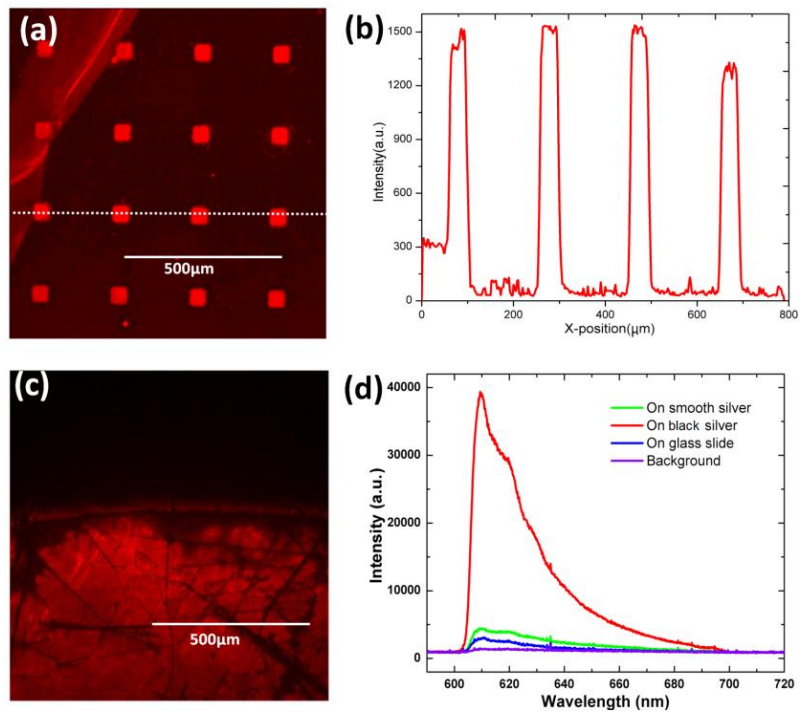

Figure 2. Fluorescence enhancement of R6g on black silver substrate. (a) Pseudo-colored fluorescent intensity image on square array patterned black silver. (b) Cross section intensity profile along the white dashed line on (a). (c) Pseudo-colored fluorescent intensity image on the edge of a R6g drop deposited on an uniform black silver substrate. (d) Fluorescence spectra of R6g on smooth silver, regular glass slide and uniform black silver substrate.

To demonstrate fluorescence enhancement on the black silver substrate, we deposit R6g solution with the concentration of 10 μM on both black silver, smooth silver and glass slide, wait until dry and excite the fluorescence with green light (550nm center wavelength). The image is taken with a microscope objective lens with 20x magnification and numerical aperture (NA) of 0.5. **Fig. 2a** is the intensity image taken on the square array patterned black silver substrate. Obviously, the intensity on the black silver square region is much higher than the surrounding smooth silver region, which is also illustrated in **Fig. 2b**, the intensity profile across the white dashed line on Fig. 2a. **Fig. 2c** is the fluorescence intensity image on the edge of a R6g drop stain on uniform non-patterned black silver, in which the red region is covered by R6g and evidently the uniform molecule deposition ensures fair intensity measurement. From **Fig. 2d**, the comparison of fluorescence emission spectra over the entire microscopic field of view (400μmX400μm) taken on uniform black silver, smooth silver and regular glass slide, all captured with the integration time of 5 seconds, we see the emission of R6g is much stronger on black silver than on smooth silver or glass slide. By subtracting the background from all the fluorescence spectrums and dividing the area under spectra curve from 600nm to 700nm on black silver by that on smooth silver and glass slide, we obtain the fluorescence enhancement on the black silver for 15 times with respect to smooth silver and nearly 30 times with respect to glass slide. It is worth noting that the fluorescence enhancement on the photon trapping black silver substrate is an interesting observation. For most plasmonic metal nanostructures like silver or gold nanoparticle enhanced fluorescence, the fluorescence emission photon is not likely to be trapped, therefore almost all emission photons may be acquired in imaging process. However due to the highly efficient photon trapping property of black silver the fluorescence emission photons from R6g molecules may be mostly trapped within the nanocone forest without being detected. Even with such potential loss, we still observed 30 times fluorescence enhancement. We provide two possible explanations: the first one is that the fluorescence emission photons from the R6g molecules are mostly converted into plasmons and later re-emitted into free space through the plasmon scattering which was accounted for the 10% optical "reflectance" measured from the black silver substrate; the second explanation is that the cavity mode plasmon resonance and localized electromagnetic field in the silver coated nanocone forest are extremely strong and can excite very high fluorescence emissions for which even a small portion of fluorescence emission photons escaping to the free space has much higher intensity than the fluorescence intensity on smooth silver or glass slide. Further theoretical investigation is underway.

As a surface plasmon enhanced phenomena, SERS can be modeled with electromagnetic field theory and the enhancement factor G can be estimated as the fourth power of the electrical field amplitude E, or G is proportional to $E^4$.[15] With finite element modeling method implemented in the software COMSOL, we simulate the two-dimensional electric field distribution around silver covered silicon nanocone structure, shown in **Fig. 3a**. To approximate the structure shown in Fig.1c, we draw an array of close-packed silicon nanocones with the height of 500nm, width of 180nm at the bottom and the period spacing of 180nm. Although the real nanostructures do not have the perfect periodicity as drawn in the simulation, the simulation for electric field enhancement should still represent the case in the actual silver coated nanocone substrate in principle. The localized electromagnetic field distribution agrees for both simulated and actual nanostructures as they share the basic nanoscale profiles and material properties. The only discrepancy between the simulated and actual cases is the discrete resonance modes for the period photonic crystal structure in simulation which do not agree with the actual pseudo randomly distributed nanocone arrays. Evidenced by the SEM image in Fig. 1c, most of the deposited silver is likely to reside on top of the nanocones and the deposition on the side wall is thinner. In the simulation model we set the side wall covered by only 15nm thick silver and a silver bead with diameter of 80nm on the tip of each nanocone. The optical constant of silver and silicon at a certain wavelength is retrieved by polynomial fitting to the data looked up in the reference handbook.[16] With the incident 785nm transverse

magnetic(TM) polarized plane wave propagating in –Y direction,[14] the excited scattering electric field is calculated. The color bar on the right of Fig.3a indicates the normalized amplitude of scattering electric field with respect to the amplitude of electric field of incident wave. We can see the scattering electric field is largely enhanced in the regions between adjacent silver beads due to plasmon coupling effect. The maximum electric field enhancement which is in the proximity of the bead surface can reach 160 times. The electric field inside the cavity sandwiched by two adjacent nanocones is also amplified especially near the valleys. Due to the unique "nanocavity"[17] profiles of the nanocone arrays, multiple plasmon resonance modes[18] in a very broad wavelength range exist and contribute to the high field enhancement.

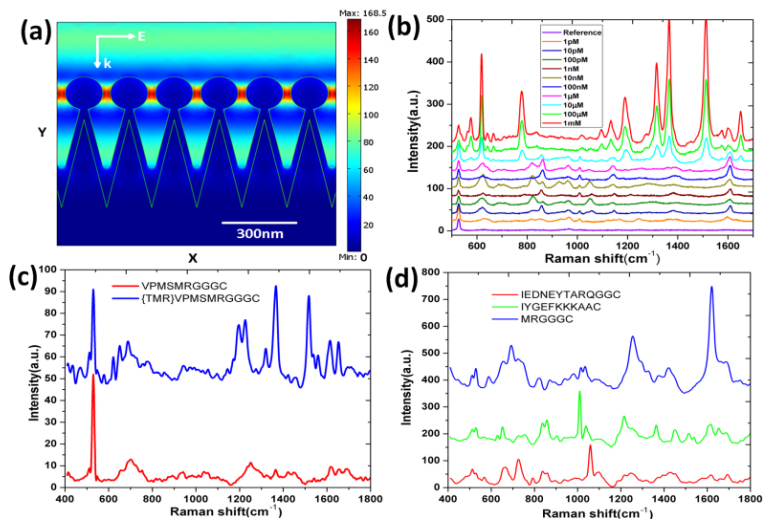

Figure 3. Raman scattering enhancement and label-free peptide SERS sensing on black silver substrate. (a) 2D FEM simulation of electric field distribution around the black silver (b) SERS spectra for R6g solution deposited on black silver with the gradient concentrations. (c) SERS spectra of 100 nM peptides of the same sequence without label and with TMR label. (d) SERS spectra of 100 nM unlabeled peptides with three different sequences.

To experimentally interrogate the surface Raman scattering enhancement of black silver substrate, we measure the Raman spectra of R6g solutions with the various concentrations from 1 mM to 1 pM diluted by 10 times between each deposited onto a black silver substrate (**Fig. 3b**) along with that of a R6g solution with the concentration of 10mM deposited on smooth silicon wafer for reference. By analyzing the intensity of the R6g characteristic peak at 1370cm$^{-1}$ in each case with a commonly used characterization method[12,19], the averaged enhancement factor is calculated as $6.38 \times 10^7$.

Other than the SERS detection of chemical molecules on black silver substrate demonstrated, we also present its application in non-labeled and labeled biomolecule detection, e.g. peptide sensing. **Fig. 3c** shows the SERS spectra of 100 nM labeled and unlabeled peptide with the same amino acid sequence VPMSMRGGGC deposited on a black silver substrate. **Fig. 3d** shows the SERS spectra of 100 nM unlabeled peptides with three different sequences, MRGGGC (blue curve), IEDNEYTARQGGC (red curve) and IYGEFKKKAAC (green curve) deposited on a black silver substrate. Three sequences show different characteristic peaks in Raman spectra which allow them to be distinguished and identified without labeling.

All SERS measurements are carried on with the same setup and configuration. A laser with wavelength of 785nm and power of 30mW is used for Raman excitation. The scattered light is collected with an objective with 10x magnification and NA = 0.28. We keep integration time of 5 seconds for capturing all the spectrums. Prior to the analysis and plotting, the fluorescence background has been removed from all the spectrums with an automated iterative polynomial fitting algorithm.[20] All peptides we use have Cysteine(C) with a thiol group for binding to silver at one end of the sequence.

In summary, we present a nanomaterial substrate, which we call black silver, produced by depositing silver on black silicon fabricated with plasma etching process and we demonstrated the broadband strong enhancement effects for multiple optical properties including absorption, fluorescence and Raman scattering from UV to NIR wavelength range. In the end, the label-free peptide and chemical sensing on black silver substrates is experimentally demonstrated.

**Acknowledgement**

This work is in part supported by NSF grant ECCS 10-28568 and Illinois ECE startup fund. The author would like thank Dr. Larry Millet for the assistance of fluorescence spectroscopy measurement.